\documentclass{aastex}

 \usepackage{CJK}

\shorttitle{New Member of the Inner Oort Cloud from NGVS}
\shortauthors{Chen et al.}

\begin{document}
\begin{CJK*}{UTF8}{bkai}

\title{Discovery of a New Member of the Inner Oort Cloud from The Next Generation Virgo Cluster Survey}

\author{Ying-Tung Chen (陳英同)}
\affil{Institute of Astronomy, National Central University, No. 300, Jhongda Rd, Jhongli City, Taoyuan County 32001, Taiwan}
\email{charles@astro.ncu.edu.tw}

\and

\author{J. J. Kavelaars, Stephen Gwyn, Laura Ferrarese, Patrick C\^ot\'e}
\affil{Herzberg Institute of Astrophysics, National Research Council of Canada, Victoria, BC V9E 2E7, Canada}

\and
\author{Andr\'es Jord\'an, Vincent Suc}
\affil{Departamento de Astronom\'ica y Astrof\'isica, Pontificia Universidad Cat\'olica de Chile, Av. Vicu\~na Mackenna 4860, Macul 7820436, Santiago, Chile}

\and

\author{Jean-Charles Cuillandre}
\affil{Canada-France-Hawaii Telescope Corporation, 65-1238 Mamalahoa Hwy., Kamuela, HI 96743, USA}

\and

\author{Wing-Huen Ip (葉永烜)}
\affil{Institute of Astronomy, National Central University, No. 300, Jhongda Rd, Jhongli City, Taoyuan County 32001, Taiwan}

\begin{abstract}
We report the discovery of 2010 GB$_{174}$, a likely new member of the Inner Oort Cloud (IOC). 2010 GB$_{174}$ is one of 91 Trans Neptunian Objects (TNOs) and Centaurs discovered in a 76~deg$^2$ contiguous region imaged as part of  the Next Generation Virgo Cluster Survey (NGVS) --- a moderate ecliptic latitude survey reaching a mean limiting magnitude of $g^\prime \simeq 25.5$  --- using MegaPrime on the 3.6m Canada France Hawaii Telescope. 2010 GB$_{174}$ is found to have an orbit with semi-major axis $a\simeq350.8$~AU, inclination $i \simeq 21.6^\circ$ and pericentre $q\sim48.5$~AU. This is the second largest perihelion distance among known solar system objects. 
Based on the sky coverage and depth of the NGVS, we estimate the number of IOC members with sizes larger than 300 km ($H_V \le 6.2$ mag) to be $\simeq 11\,000$. A comparison of the detection rate from the NGVS and the PDSSS (a characterized survey that `re-discovered' the IOC object Sedna) gives, for an assumed a power-law LF for IOC objects, a slope of $\alpha \simeq 0.7 \pm 0.2$, with only two detections in this region this slope estimate is highly uncertain. 
\end{abstract}


\slugcomment{Submitted to Astrophysical Journal Letters}

\section{Introduction} 
The exact delineation of the Inner and Outer Oort cloud is not well known at the present time. A distinction between the two regions is supported by considerations of the evolution of cometary orbits \citep{Hills81, Duncan87}. Historically, the boundary between the two clouds has defined by the minimum semi-major axis a comet must have to be sufficiently perturbed by Galactic tides or stellar encounters to enter the inner solar system.  The Inner Oort Cloud  (IOC) region, however, may also be a significant source of comets \citep{kai09, fou13}.  A lower bound on the inner edge of the IOC is the outer edge of the ``scattered disk", whose dynamical evolution is influenced by interactions with Neptune; formation models had traditionally placed the inner edge at much greater distances (i.e., 1000s of AU).  Objects with semi-major axis $a \gtrsim 200$~AU and perihelion $q \gtrsim44$ AU do not experience substantial perturbations of their orbital elements from the giant planets,  as those elements remain largely unchanged during long-duration numerical simulations \citep{gla02, lyk07b}. Being decoupled from strong interactions with Neptune resonances, objects in this region are therefore possible members of the IOC.  
\citet{lev97} showed that objects with $q >$ 40~AU have an extremely low probability of exiting the Kuiper belt to become Jupiter family comets. 
Numerical integrations of outer solar system objects indicate that they do not scatter onto high perihelion ($q > 40$~AU) orbits via interactions with the known planets alone \citep{gla02, eme03, gom05}, and objects with $q > 40$~AU do not have significant encounters with Neptune over the age of the solar system \citep{lyk07a}.
The Kozai mechanism can lift the perihelia distances by exchange of inclination and eccentricity, but it does not drive objects with smaller inclination ( $i~\leq$ 20$^\circ$ ) and large perihelion distance ($q > a/27.3 + 33.3$~AU) \citep{gal12}.  \citet{lyk07b} found that  resonance sticking  can lift the perihelia of scattered disk objects, but this mechanism is unimportant for orbits with $a > 250$~AU.    Thus, somewhere beyond 200~AU, a transition to the meta-stable IOC region appears to occur.

The large perihelion distances and high eccentricities of IOC objects must  have originated via perihelion lifting mechanisms triggered by external perturbations, which are unlikely to occur within the Sun's current galactic environment.
Proposed scenarios include 
interactions with a planet-sized body in the outer Oort cloud \citep{lyk08, gla06, gom06, mat05, mat11};
the passage of a $\sim$ solar mass star within a few hundred AUs  \citep{mor04};
solar migration within the Milky Way \citep{kai11};
stellar encounters that took place while the Sun was still a member of its natal star cluster  \citep{bra06, bra07, bra12, kai08}; and
the capture of extrasolar planetesimals from low-mass stars during the early evolution of the solar system \citep{ken04, mor04}. Since discriminating between these scenarios is important for understanding the early evolution of the solar system, new observational constraints on the orbits of IOC objects are of obvious interest.

Observationally, the notion of an IOC gained considerable traction with the discovery of 2000 CR$_{105}$ \citep{gla02}, the first discovered member of the IOC. Two years later, the discovery of Sedna  \citep{Brown04} --- $q \sim76$~AU and large eccentricity --- dramatically extended the size of the known solar system. Sedna's detection was followed by the discovery of 2004 VN$_{112}$ \citep{Becker08}, $q \sim 47$ AU and $e \sim 0.86$. These objects (see summary in Table~\ref{tb1}) point to the existence of an extended reservoir of material beyond the scattered disk of Trans Neptunian Objects (TNOs), probably coinciding with the inner edge of the Oort Cloud \citep[eg.][]{Gomes10}.  

We report the discovery of a likely new member of the IOC, 2010 GB$_{174}$, based on Canada France Hawaii Telescope (CFHT)  MegaPrime \citep{bou03} imaging from the Next Generation Virgo Cluster Survey \citep[NGVS;][]{fer12}.

\section{Observations}

The NGVS is a deep, high spatial resolution imaging survey of 104 deg$^2$ within the Virgo cluster, using the MegaPrime imager at CFHT, a  detailed description can be found in \citet{fer12}. 
The dithering strategy adopted by the NGVS yields an  effective field of view for MegaPrime of 1 deg$^2$, sampled with a pixel scale of $0\farcs187$. 
For the $g^\prime$- and $i^\prime$-band exposures used in our TNO search, the minimum time separating exposures of the same field is 1.3 and 0.8 hours, respectively, although this interval can stretch to days, weeks, or even months. 
This results in a cadence that, although not specifically optimized for the detection of outer solar system objects, does allow for such objects to be detected.
Single NGVS images have exposure times  of 411 seconds in $i^\prime$ and 634 seconds in $g^\prime$, corresponding to point-source depths of $24.7$ and $25.5$ mag, respectively  (1.3$\sigma$), making this the deepest wide-field survey for outer solar system objects yet conducted. 

For the TNO search described in this paper, individual frames were first bias corrected, dark subtracted and flat fielded, using the standard {\tt Elixir} pipeline adopted for all MegaPrime images \citep{mag04}. 
The astrometry and photometry were calibrated on the SDSS system using the {\tt MEGAPIPE} pipeline \citep{gwy08}. The resulting five calibrated images (for each field in each of the $g^\prime$ and $i^\prime$ filters) formed the input observations for our TNO search process. 

Moving sources were detected using image differencing. 
First, individual frames in a common filter
were stacked to form a high-S/N template.  The template  was subtracted from  individual frames using the {\tt SWarp} package \citep{bert02}.
Finally, {\tt SExtractor} \citep{bert96} was used to generate a list of all detections in the resulting, differenced images.  
Sources that exhibited high peak flux contrasts relative to normal stellar PSFs were rejected as likely cosmic ray events; remaining detections larger than 1.3$\sigma$ were kept as our non-stationary sources. 
Detections on two separate frames taken during the course of a single night --- and whose separation was consistent with motions between $0\farcs735$ and $7\farcs35$ per hour, corresponding to objects on circular orbits located between 20 AU and 200~AU --- were linked to form a pair. 
For a detection to be considered real, we required that at least four individual measurements taken in a single lunation could be linked together.
The pairs were linked via trial fitting to Kuiper belt orbits using the \citet{ber00} software (BK hereafter).
Trial links that exhibited large fit residuals ($>0\farcs5$ per measure) were rejected and smaller fit residual trial links were further examined to confirm candidate detections. 
The candidate detections that passed a visual inspection were then accepted as {\it bona fide} TNOs.

\subsection{Detection Efficiency}

We determined our detection efficiency by ``planting'' (adding) artificial sources in each frame.  
A detailed description of our approach can be found in \citet{Hu11}. 
Approximately 10\,000 artificial sources per square degree were added to the images.
Artificial objects were considered detected if a linked pair was recovered in images following the same cadence as the original dataset.
Using these artificial sources we determine a survey limit of $g^\prime \simeq 25.5$, which was sensitive to Pluto-sized objects out to 400 AU.

\subsection{Orbit Tracking}

Owing to the survey cadence, NGVS allows internal recovery observations  for most of discovered TNOs, as long as the TNO remains within NGVS footprint. 
Additionally, public archives from a variety of telescopes can be exploited in the off chance that the object might have been recorded serendipitously.
The search for additional detections of TNOs was performed using the Solar System Object Search \citep[SSOS;][]{Gwyn12}. Detected sources were paired (based on a motion requirement consistent with TNO orbits) and these pairs were then kept as possible new observations of the TNO of interest. The goodness of fit for the orbit was determined using 
{\it fit\_radec} (BK software), and pairs for which the fit residuals were $<0\farcs5$ were visually inspected.
If  both members of the source pair passed visual inspection then the new pair was accepted as a valid recovery of original TNO. 
Using this approach of recovery, the arcs of many detected TNOs were extended to months and years in length.

Of course, several TNOs moved outside NGVS footprint or dimmed below NGVS limit before they could be recovered. 
We therefore obtained additional recovery observations using both CFHT/MegaPrime and the  MegaCam imager \citep{mac98} on the Magellan 6.5m Clay telescope.
These observations, acquired in spring 2010 and 2011 semesters, targeted the predicted locations of the NGVS TNOs.
Additional follow-up imaging, acquired as part of the NGVS, will be completed by June 2013.

\begin{figure}
\plotone{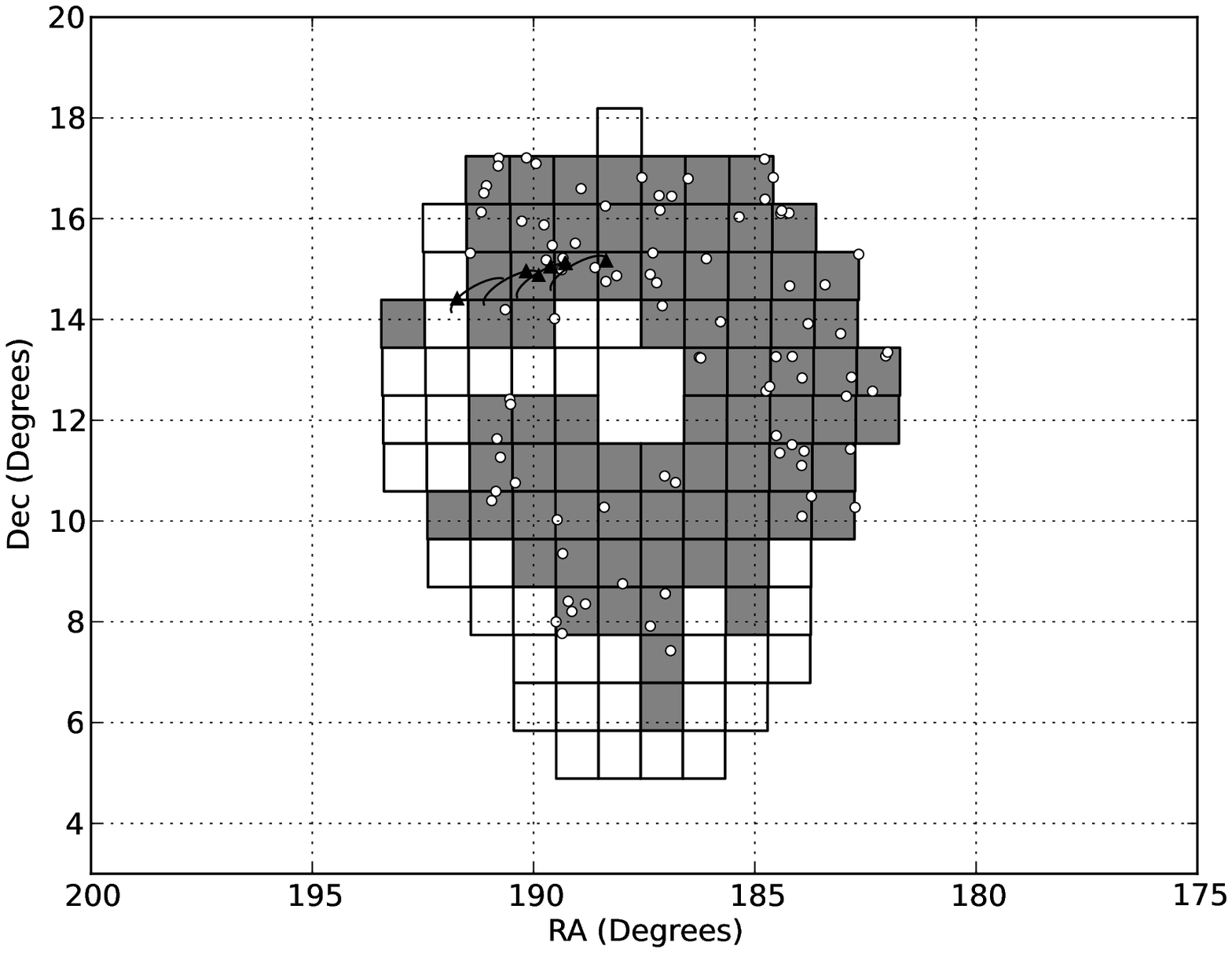}
\caption{NGVS coverage in the $g^\prime$ band as of 2012 plotted on the J2000 sky. The open circles are the 91 objects discovered in the search fields (plotted as gray squares). Individual observations of 2010 GB$_{174}$ are shown by the triangles. Predicted and observable positions for 2010 GB$_{174}$ are indicated by curves (moving from west to east during the period 2009 to 2012). \label{fig1}}

\end{figure}

\begin{figure}
\plotone{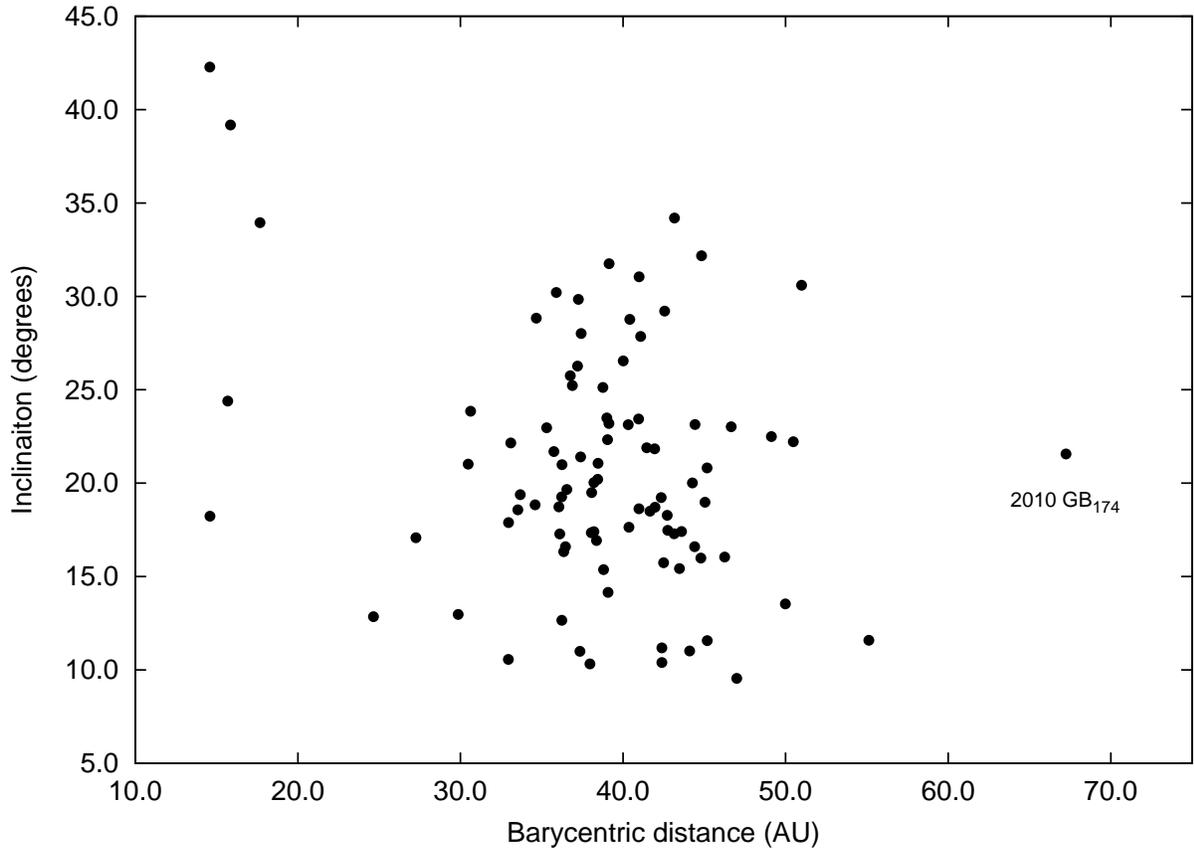}

\caption{Plot of inclination versus barycentric distance for all 91 discoveries from the NGVS. 2010 GB$_{174}$ is the outermost object in this figure, at a barycentric distance of $\sim 67.2$ AU.\label{fig2}}
\end{figure}

Based on the procedure outlined above, a total of 91 new TNOs and Centaurs have been discovered (see Figure \ref{fig1}). 
Once the recovery observations are used, 68 of these 91 objects have a semi-major axis uncertainty of less than 5\%, and of these, 65 have more than 60 days of arc length.
All sources whose orbital parameters at detection were consistent with a $q>45$~AU orbit (i.e., those objects discovered 45~AU from the Sun) were tracked until their BK fitting result excluded $q > 45$~AU solutions. 
The distance at detection versus orbital inclination for all our NGVS discoveries is shown in Figure \ref{fig2}. 
Full details on the complete samples will be presented in Chen et al. (in preparation). For the remainder of this paper, we shall focus on a single object found to have $q \sim 48$~AU and $a \sim 350$~AU.

\section{Analysis and Results}

\subsection{Orbit and Magnitude}

\begin{figure}[htp]
\begin{tabular}{cccc}
    \includegraphics[width=0.25\textwidth]{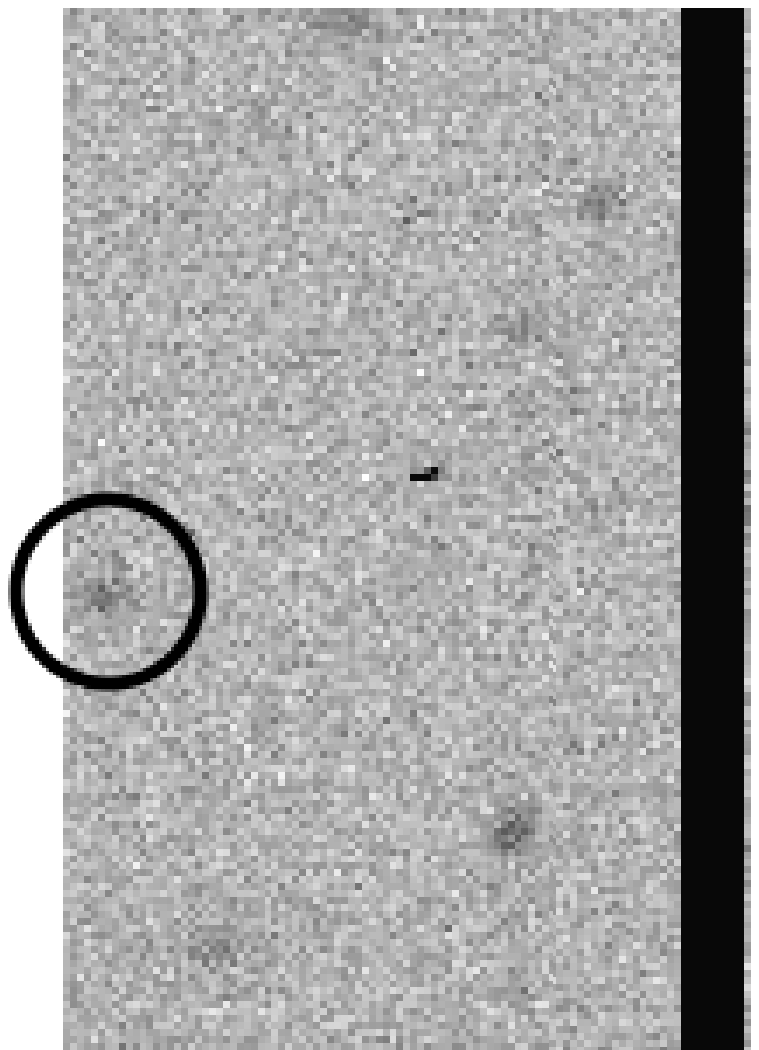}&
    \includegraphics[width=0.25\textwidth]{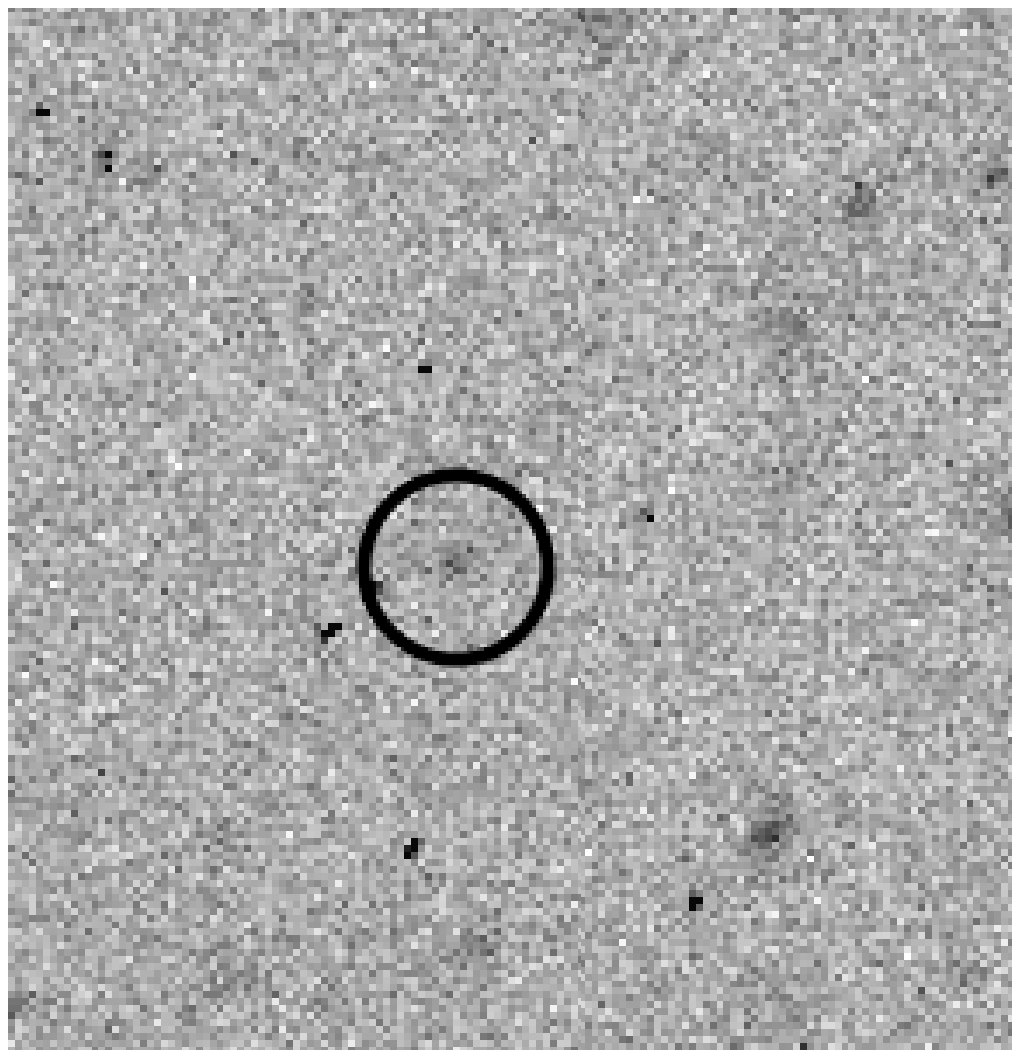}&
    \includegraphics[width=0.25\textwidth]{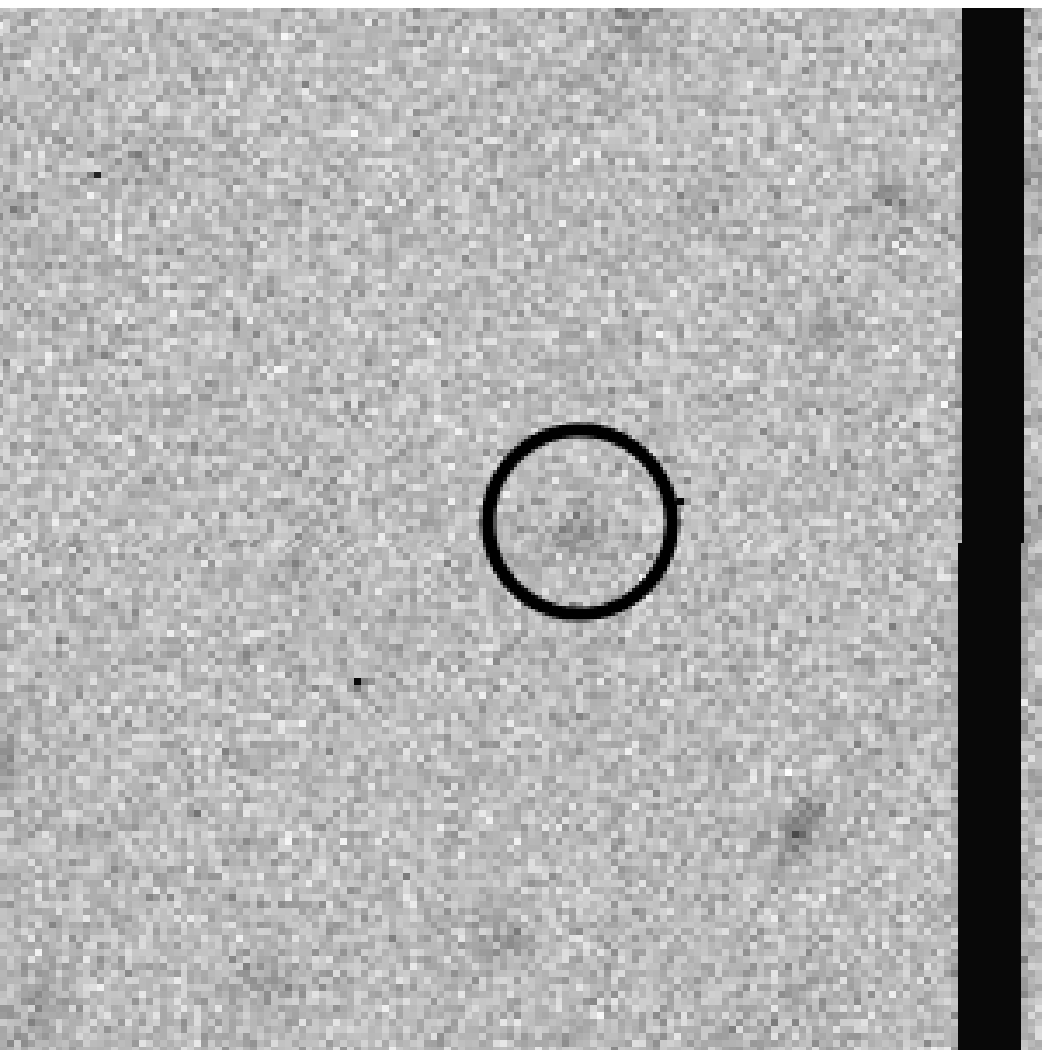}&
    \includegraphics[width=0.25\textwidth]{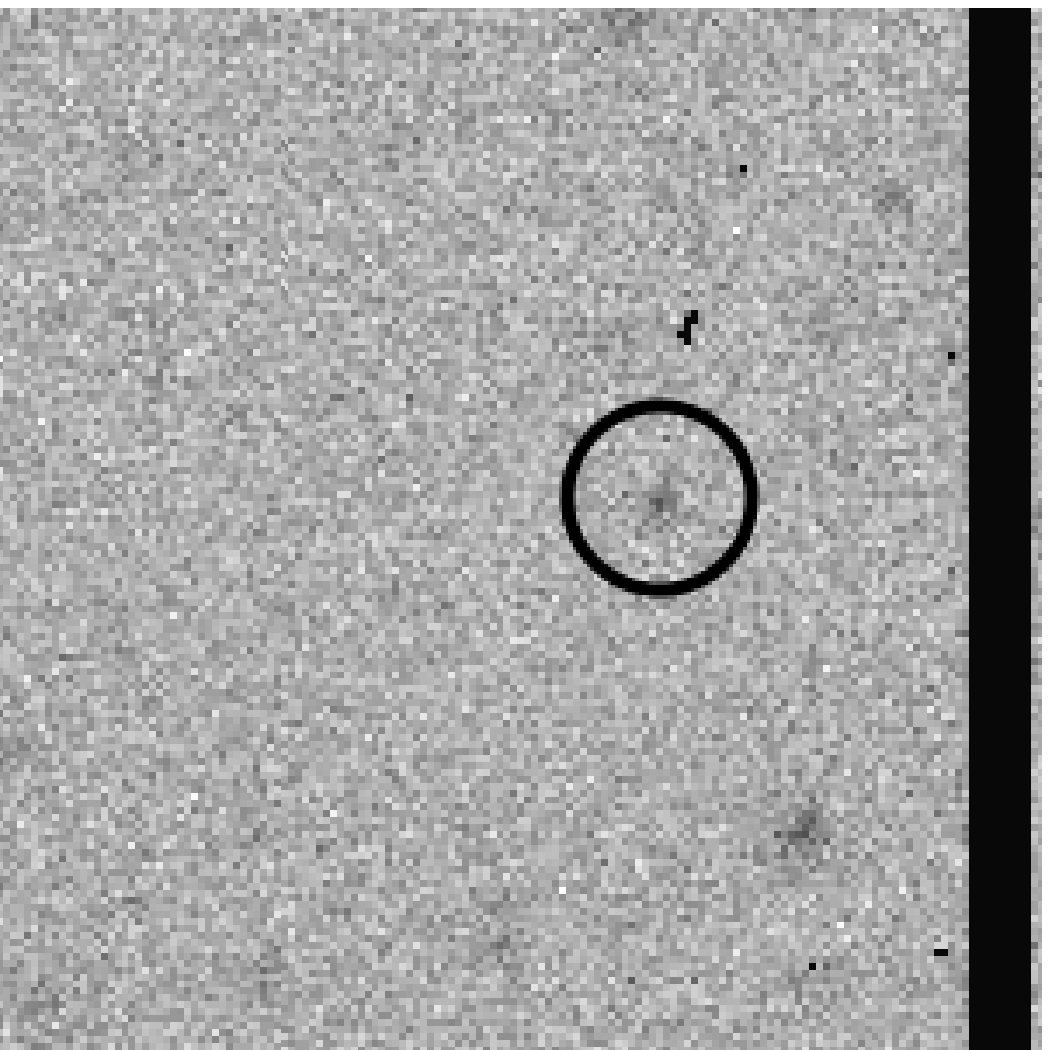}\\
\end{tabular}
\caption{Discovery images of 2010 GB$_{174}$ obtained with the MegaPrime instrument on CFHT. From left to right, the 634 sec exposures were obtained on 2010 April 13 at 7:03, 8:26, 9:49, and 11:11 (UT), respectively. The object moves $7\farcs2$ over a period of 4.1 hours.}\label{fig3}
\end{figure}

2010 GB$_{174}$, which was discovered in the 2010 NGVS data (Figure \ref{fig3}), is the slowest moving  ($1\farcs76$~hr${^{-1}}$) object detected in our survey, consistent with a location beyond 60 AU from the Sun.
Its $g^\prime$ magnitude is found to be 25.0$\pm$0.2 (absolute magnitude H$_{g^\prime} \simeq  6.7$), with no evidence for large ($\pm 0.3$ mag) amplitude variability. 
The object was recovered in NGVS images taken one month after the discovery epoch, suggesting an orbit with $a \sim 100$~AU and $e \sim 0.3$. 
Follow-up Magellan images taken in 2010 and 2011 --- combined with pre-discovery images from the NGVS taken in 2009 --- provide three oppositions of observations at a variety of elongations, resulting in a robust orbit determination. 
The best-fit orbit has a semi-major axis of $a = 350\pm4 $~AU, an eccentricity of $e  = 0.862 \pm 0.002$, and an inclination of $i = 21\fdg563\pm0\fdg001$, detected at a barycentric distance of $67.24$~AU.
Given these orbital parameters, 2010 GB$_{174}$ occupies a region between $48.5$~AU and $654.9$~AU from the Sun: i.e., it is firmly placed between  the outer boundary of the Kuiper belt  and the Oort Cloud.

To test for long-term stability, we performed numerical simulations using {\tt MERCURY} \citep{Chambers99}.  
We find that this object and clones consistent with its current orbital uncertainty  are very stable (i.e., $\delta a  \pm 15$~AU and $\delta e \pm 0.03$) over periods in excess of 4 Gyr.   
None of the orbital elements exhibit variations that would indicate resonance trapping or sticking.
The lack of resonance sticking or other resonant-like evolution is consistent with results from modelling of this region of orbital parameter space that suggest objects with $a > 250$~AU do not undergo significant orbital evolution \citep{lyk07b}.
The lack of orbital evolution, along with the large pericentre and high orbital eccentricity, lead us to conclude that 2010 GB$_{174}$, the second largest perihelion solar system object, is a new member of the IOC.

\subsection{Luminosity Function of the IOC} \label{lumi}
Given our one IOC detection, and assuming that this object is drawn from the same population as Sedna, we can make a rough estimate of the luminosity function of the IOC.  
First, we note that Sedna was (re)discovered in the Palomar Distant Solar System Survey (PDSSS) \citep{Schwamb10}, a survey that covered 12\,000 deg$^2$ to an $R$-band limit of $\sim 21.3$ mag. No other IOC objects were reported in this survey.  By contrast,
the NGVS covered 76 deg$^2$ with images acquired in a sequence that enabled the detection of moving objects.  These images were searched for moving objects to a limiting magnitude of $R$  = 24.5\footnote{Using $(g^\prime-r^\prime)$ = 0.8 from \citet{she10}, and $V = g^\prime-0.55(g^\prime-r^\prime)-0.03$ and $(V-R) = 0.59(g^\prime-r^\prime)+0.11$ from \citet{Smith02}, to obtain $R \sim g^\prime - 1.0$.}  and, like the PDSSS survey, yielded a single IOC object. 
Comparing the sky coverage and flux limits of the PDSSS and NGVS surveys, and assuming that the IOC luminosity function can be described by a standard power-law, $\Sigma(m) = 10^{\alpha(m-m_0)}$, yields a slope of 
$$ \alpha = \log_{10}(12000/76)/(24.5-21.3) \simeq 0.7 \pm 0.2$$, where the slope uncertainly only takes into account Poisson counting uncertainties.

Being based on only two detections this represents a hint at the underlying luminosity function.  The luminosity function of large objects (like Sedna) are known to differ significantly from that of objects with $H \gtrsim 4$, perhaps due to size dependent albedo variations \citep[eg.][]{2008ssbn.book..335B}.  Our estimated luminosity function slope, while consistent with that reported for other outer solar system populations  
\citep[e.g.,][]{fra10}, spans an H range where considerable structure in the LF is likely to exist and so should only be taken as a first  indication of the underlying distribution.   
More detections of distant detached/IOC objects are clearly needed before firm conclusions on the slope and shape of the luminosity function can be made.

A power-law slope of $\alpha = 0.7$ is consistent with the fact that no IOC objects were detected by the CFEPS  project \citep{petit11}, which surveyed an area of $\sim 200$ deg$^2$ to a limiting magnitude of $g^\prime \sim 23.5$ ($R \sim 22.5$). A power-law with $ \alpha \simeq 0.7$  and our single detection at $V \sim 24.5$ in 76 deg$^2$ predicts $\sim$ 0.1 IOC detections in the CFEPS survey. In other words, one detection would be expected only if CFEPS were to survey $\sim 1800$ deg$^{2}$.
Based on these rough calculations, the very small numbers of IOC objects reported in the literature to date is not surprising.

\subsection{Population Estimates} \label{pop_est}

The very fact that 2010 GB$_{174}$ was detected at all implies a large IOC population. 
Based on its measured orbit, 2010 GB$_{174}$ is expected to be brighter than the NGVS detection limit for only $\sim 1.8\%$ of its orbital period. Moreover, given the orbital inclination, 2010 GB$_{174}$ can occupy $\sim 14\,400$ deg$^2$ of sky. 
Despite these strong selection effects,  2010 GB$_{174}$ was detected in a survey covering just 76 deg$^2$.
Combining these constraints implies a total IOC population of $N_{\rm IOC} \sim 14\,400/76/0.018 \sim 11\,000$ objects with $H_V < 6.2$\footnote{$H_V = H_{g^\prime} -0.5 = 6.7 - 0.5 = 6.2$, using the same references in \S\ref{lumi}.}.
This population estimate is a lower-limit for the IOC population as our survey is significantly incomplete for peri-centres larger than 100AU where a significant IOC population is likely to exist. 

Based on the fact that Sedna was the only IOC object detected in a 12\,000 deg$^2$ area, and assuming a luminosity function with slope $\alpha =0.58$, \cite{sch09} estimated a total IOC population of $N_{\rm IOC} \approx 40$ for $H_V<1.6$. 
Combining these different population estimates and assuming a power-law luminosity function, implies $\alpha \sim \log_{10}(11\,000/40)/(6.2-1.6) \sim 0.53$, a slope that is consistent with the one derived in \S\ref{lumi} derived by simply comparing the depths and areal coverage of the NGVS and PDSSS.  The similarity between this slope estimate and the preceding estimate based on the brightness at detection gives some support to notion that 2010 GB$_{174}$ and Sedna are, indeed, samples from the same underlying population even though they have substantially different peri-centres. 

\section{Discussion}


An examination of the population ratio of the ``scattered" disk to the Inner and Outer Oort clouds may be a diagnostic of formation models.  
Models of close stellar passages tend to produce large numbers of objects with $q > 40$~AU and $a > 200$~AU \citep[e.g.,][]{kai08}. 
These models result in population ratios of IOC to scattered disk objects in the range $\sim$ 8--16, with no strong dependence on the specific stellar environment in which the Sun formed.  
We can estimate the size of the scattered population  (i.e., objects with $33 < q < 40$~AU and $a > 48$~AU) using the CFEPS Outer/detached population from \citet{petit11}. 
By scaling the CFEPS population estimate  ($N(H_g < 8) = 13\,000$ and $\alpha = 0.8$) to the H magnitude limit of our survey  ($H_g = 6.7$) 
we obtain $N(H_g <6.7) = 13\,000\times(10^{0.8\times(6.7-8)}) \sim 1200$,  a number that is a factor of 10 less than our IOC population estimate at this H limit.  
This ratio is consistent with estimates of the IOC to scattered disk ratios found in close stellar passage models.  
This agreement between our estimated ratio for the IOC to scattered disk population is in stark contrast to the disagreement between model estimates of the ratio between the Outer Oort cloud and scattered disk and observation based measurements of that ratio. 
Perhaps this disagreement originates in the process used for outer Oort cloud population estimates rather than in the solar system formation models,  as \citet{bra13} have also mentioned.

\section{Summary}

We have reported the discovery of a new detached/IOC object, 2010 GB$_{174}$, in a search for TNOs covering a 76 deg$^2$ region from the NGVS. 
The survey limiting magnitude is $g^\prime \sim 25.5$ (1.3$\sigma$), corresponding to $H_{g^\prime} = 6.7$ mag and a TNO size of 300 km (for an assumed an albedo of 8\%).
2010 GB$_{174}$ is the second most distant perihelion object known in the solar system.
Given 2010 GB$_{174}$'s orbital parameters, and accounting for the limiting magnitude of the NGVS, we estimate the total number of large perihelion, large semi-major axis objects with $H_V <  6.2$ to be $\gtrsim$ 11\,000.
Combining survey area and limiting magnitude of NGVS with PDSSS and assuming a power-law luminosity function indicates a slope of $\alpha \sim 0.7\pm 0.2$ (only account Poisson counting uncertainties). 
The IOC is a key domain for characterizing the early environment of solar system, and the discovery of 2010 GB$_{174}$ --- with its large semi-major axis and perihelion distance --- provides further evidence for an extensive IOC population. 

\acknowledgments
This work is supported by NSC101-2119-M-008-007-MY3 and the Canadian Advanced Network for Astronomical Research (CANFAR). 
We thank the NGVS team and the queued service observing operations team at CFHT for their excellent arrangement of the NGVS observations.
AJ acknowledges support from Fondecyt project 1130857, Anillo ACT-086, BASAL CATA PFB-06 and the Millennium Science Initiative, Chilean Ministry of Economy (Nucleus P10-022-F).
This research also used the facilities of the Canadian Astronomy Data Centre operated by the National Research Council of Canada with the support of the Canadian Space Agency. 
We acknowledge Matthew J. Holman for helpful conversations and Nathan Kaib for useful comments on the manuscript. Finally, we thank Robert Jedicke and Larry Denneau for help with the PS1 MOPS.

{\it Facilities:} \facility{CFHT}, \facility{Magellan:Clay}.

\clearpage

\clearpage

\clearpage

\begin{table}
\caption{Orbital Elements for All Possible Members of the Inner Oort Cloud \label{tb1}}
\begin{tabular}{lccccc}\\
\tableline
\tableline
Object              & $a$  & $q$  & $e$ & $i$ & $H_V$\\
 & (AU) & (AU) & & (deg) & (mag) \\
\tableline
Sedna                          & 542.7 & 76.2 & 0.86 & 11.9 & 1.6\tablenotemark{a}\\
2004 VN$_{112}$     & 343.3 & 47.3 & 0.86 & 25.5 & 6.4\\
2000 CR$_{105}$     & 228.8 & 44.1 & 0.81 & 22.7 & 6.1\\
\tableline
2010 GB$_{174}$                            & 350.7 & 48.5 & 0.86 & 21.5 & 6.2\tablenotemark{b}\\
\tableline
\tableline
\end{tabular}
\tablecomments{Except 2010 GB$_{174}$, all orbital elements were taken from: {\tt http://ssd.jpl.nasa.gov/sbdb.cgi }}
\tablenotetext{a}{Absolute magnitude from \citet{Schwamb10}.}
\tablenotetext{b}{$V$-band absolute magnitude converted from $g^\prime$ as described in Section~\ref{pop_est}.}

\end{table}

\end{CJK*}
\end{document}